\documentclass [12pt]{article}
\usepackage{graphicx,amssymb,amsfonts,latexsym,times,amsmath,amsthm}
\usepackage{epsfig}
\begin{document}

\title{A new view on migration processes between SIR centra: an account of
the different dynamics of host and guest }

\date{}

\maketitle

\centerline{\bf Igor Sazonov$^a$\footnote{Corresponding author. E-mail: i.sazonov@swansea.ac.uk}, Mark Kelbert$^b$ and Michael B.\ Gravenor$^c$
}

\centerline{$^a$ College of Engineering, Swansea University, Singleton Park, SA2 8PP, U.K.}

\centerline{$^b$ Department of Mathematics, Swansea University} 

\centerline{$^b$ Institute of Life Science, School of Medicine, Swansea University} 


\begin{abstract}
We study an epidemic propagation between $M$ population centra.
The novelty of the model is in analyzing the migration of host (remaining in the same
centre) and guest (migrated to another centre) populations separately.
Even in the simplest case $M=2$, this modification is justified because it gives a more realistic description of migration processes. This becomes evident in a purely migration model with vanishing epidemic parameters.
It is important to account for a certain number of guest susceptible present in non-host cenrta because
these susceptible may be infected and return to the host node as infectives.
The flux of such infectives is not negligible and is comparable with the flux of host infectives migrated to
other centra, because the return rate of a guest individual will, by
nature, tend to be high.
It is shown that taking account of both
fluxes of infectives noticeably increases the speed of epidemic spread in a 1D lattice of identical SIR centra.
\end{abstract}

\noindent {\bf Key words:} spatial epidemic models; migration dynamics; ourbreak time

\noindent {\bf AMS subject classification:} 92D30, 91D25

\section{Introduction}

The classical SIR model is one of the simplest models which describes
qualitatively a typical directly transmitted disease outbreak in a populated
center, and remains the building block for many, more complicated applied
epidemic models. The population is assumed to consist of three components:
susceptible (S), infected (I) and removed (R).

Models of coupled epidemic centra are of particular interest because they
describe epidemic spread through network of populated centra, and hence the
overall population is not treated as a homogenous system. This is a subject of
intensive research, we mention here just a few recent publication
\cite{BBCS2012, BG1995, GBK2001, GKGB1998, BGM2007} not trying to provide an extensive bibliography.
The old scenario, known from the middle ages, when the disease propagates
locally from a village to the neighboring villages is replaced now by
almost instantaneous propagation around the globe. This phenomena was analyzed
in a many papers (see, e.g., \cite{CV2007, CV2008}). In particular, it was observed that on
heterogeneous networks an increase in the movement of population may decrease
the size of the epidemic at the steady state, although it increases the chances
of outbreak. This motivates a detailed analysis of migration in inhomogeneous populations.

The coupling
between nodes of such a network is mainly caused by migration processes of
infectives. There are several models describing such coupling (see \cite%
{Murray,DG}), for example, in \cite{Arino2007} the influence of various
parameters on the spacial and temporal spread of the disease is studied
numerically, with particular focus on the role of quarantine in the form of
travel restrictions. In \cite{SKG08,SKG11a}, the so-called diffusion like
model is proposed and studied in the framework of a fast migration time
approximation. Note that the model in \cite{Arino2007} is a particular case
of the diffusion model when the migration time tends to infinity but the
coupling coefficient introduced in \cite{SKG08} tends to zero.

In all these models the guest population is completely mixed with the host
one, so their dynamics is indistinguishable. Nevertheless, a more detailed
consideration suggests that while the epidemic dynamics is the same, the
migration dynamics should be different, especially if considered as part of
a discrete randomized model approach (cf.\ \cite{KSG11b,SKG11c}).

In the paper we start with consideration of the simplest network of only two interacting
epidemic SIR centra and study in detail the migration processes and their
influence on the population dynamics. Moreover, our interest in the model is motivated
by the fact that it serves as a hydrodynamic approximation of a natural
Markov process describing the stochastic dynamics of the system (cf. \cite{SK2008}).
This topic will be explored more fully in a subsequent paper.

To examine the migration model we first consider here the case when epidemic
parameters are temporally switched off. The study of migration in isolation provides a simple tractable model and allows us to specify the parameters in a consistent way. Equally important, this analysis reveals that many models used in the literature (see eg \cite{DG,Wang04}) are unstable in the limit of vanishing infection. Other ones (see eg \cite{Wang03,Arino05}) remain stable but lead to non-realistic results.


Note that even an isolated SIR model cannot be integrated explicitly,
therefore a suitable approximation is required to avoid numerical
integration and to obtain practical formulas for outbreak time, fade-out
time and other parameters. In our previous works \cite{SKG08,SKG11a,SKG11c}
the so-called small initial contagion (SIC) approximation was proposed,
based on the assumption that an outbreak in every population center is
caused by relatively small number of initially infectives. This
approximation is appropriate when the model is applied to strongly populated
centra like urban centra (i.e.\ in the situation when the reaction-diffusion model is not
accurate).

In the paper we also show how the model can be generalized on the general network of epidemic centra (see Section 7). As an example a characteristic
equation for the travelling wave in a chain of the population centers is derived and its numerical solution is plotted and analyzed.

\section{Governing equations}

\label{sec:eqs}

Consider two populated nodes, 1 and 2, with populations $N_{1}$ and $N_{2}$,
respectively. Let $S_{n}(t)$, $I_{n}(t)$, $R_{n}(t)$ be the numbers of host
susceptibles, infectives and removed, respectively, in node~$n$ at time $t$.
Let $S_{mn}(t)$, $I_{mn}(t)$, $R_{mn}(t)$ be numbers of guest susceptibles,
infectives and removed, respectively, in node~$n$ migrated from node $m$ at
time $t$. Removed populations $R_{n}$, $R_{nm}$ do not affect dynamics of
all others in the framework of the standard SIR model, and we omit them from
consideration here. Then, two SIR centers (nodes) interacting due to the
migration of individuals between them are described by the following model:
the dynamics of hosts in node $n$ obeys the ODEs
\begin{eqnarray}
\dot{S}_{n} &=&-\beta _{n}S_{n}(I_{n}+I_{mn})\hphantom{{}-\alpha_{n}I_{n}}%
\quad {}-\dot{S}_{n\rightarrow m}+\dot{S}_{n\leftarrow m}  \label{eq:Sn} \\
\dot{I}_{n} &=&\hphantom{-}\beta _{n}S_{n}(I_{n}+I_{mn})-\alpha
_{n}I_{n}\quad {}-\dot{I}_{n\rightarrow m}+\dot{I}_{n\leftarrow m}
\label{eq:In}
\end{eqnarray}%
where $n=1,2$, $m=2,1$; and dot denotes the time derivative. Here the term $%
\beta _{n}S_{n}(I_{n}+I_{mn})$ appears due to infectives $I_{mn}$ migrated
from node $m$ and contributing to the total disease transmission process.
Terms $\dot{S}_{n\rightarrow m}$ and $\dot{I}_{n\rightarrow m}$ describe
migration fluxes (rates) from node $n$ to node $m$ for susceptibles and
infectives, respectively. Terms $\dot{S}_{n\leftarrow m}$ and $\dot{I}%
_{n\leftarrow m}$ describe return migration fluxes (rates) to node $n$ for
guest individuals in node $m$. We specify these below.

The dynamics of guests in node $n$ temporally arriving from node $m$ can be
described by analogous ODEs
\begin{eqnarray}
\dot{S}_{mn} &=&-\beta _{n}S_{mn}(I_{n}+I_{mn})\hphantom{{}-\alpha_n
\hat{I}_{nm}}\quad {}+\dot{S}_{m\rightarrow n}-\dot{S}_{m\leftarrow n}
\label{eq:Snm} \\
\dot{I}_{mn} &=&\hphantom{-}\beta _{n}S_{nm}(I_{n}+I_{mn})-\alpha
_{n}I_{mn}\quad {}+\dot{I}_{m\rightarrow n}-\dot{I}_{m\leftarrow n}
\label{eq:Inm}
\end{eqnarray}

We assume the migration rate is proportional to the population size in the
node from which they emigrate. So, we approximate the fluxes as%
\begin{equation}
\begin{array}{lllllll}
\dot{S}_{n\rightarrow m} & = & \gamma _{nm}^{S}S_{n}, &  & \dot{I}%
_{n\rightarrow m} & = & \gamma _{nm}^{I}I_{n}, \\
\dot{S}_{n\leftarrow m} & = & \delta _{nm}^{S}S_{nm}, &  & \dot{I}%
_{n\leftarrow m} & = & \delta _{nm}^{I}I_{nm}%
\end{array}
\label{eq:mig:rates}
\end{equation}%
where $\gamma $'s and $\delta $'s are the forward and backward migration
coefficients, respectively.

Our interest in the dynamical equations
presented above is motivated by the fact that they serve as a hydrodynamic
approximation of a Markov process model. In this context, $\gamma $'s can be
associated with the transition rate for a host individual to migrate to
another node in a unit of time, and $\delta $'s---with the transition rate
for a guest individual to return to the host node.

Clearly, average return
rates should be higher: $\gamma _{nm}^{S}<\delta _{nm}^{S},\gamma
_{nm}^{I}<\delta _{nm}^{I}$, otherwise an individual would spend most of the
time out of the home center.

Substituting (\ref{eq:mig:rates}) into (\ref{eq:Sn})--(\ref{eq:In}) and (\ref%
{eq:Snm})--(\ref{eq:Inm}) yields a closed system of ODEs: for the hosts in
  node~$n$
\begin{eqnarray}
\dot{S}_{n} &=&-\beta _{n}S_{n}(I_{n}+I_{mn})\hphantom{{}-\alpha_nI_n}\quad
{}-\gamma _{nm}^{S}S_{n}+\delta _{nm}^{S}S_{nm}  \label{ODE:Sn} \\
\dot{I}_{n} &=&\hphantom{-}\beta _{n}S_{n}(I_{n}+I_{mn})-\alpha
_{n}I_{n}\quad {}-\gamma _{nm}^{I}I_{n}+\delta _{nm}^{I}I_{nm}
\label{ODE:In}
\end{eqnarray}%
and for the guests migrated from node $m$ into node $n$%
\begin{eqnarray}
\dot{S}_{mn} &=&-\beta _{n}S_{mn}(I_{n}+I_{mn})\hphantom{{}-\alpha_n%
\hat{I}_{nm}}\quad {}+\gamma _{mn}^{S}S_{m}-\delta _{mn}^{S}S_{mn}
\label{ODE:Snm} \\
\dot{I}_{mn} &=&\hphantom{-}\beta _{n}S_{mn}(I_{n}+I_{mn})-\alpha
_{n}I_{mn}\quad {}+\gamma _{mn}^{I}I_{m}-\delta _{mn}^{I}I_{mn}.
\label{ODE:Inm}
\end{eqnarray}%
Evidently, the dynamics of hosts and guests are different.

Typical initial conditions for epidemiological problem describe a number of
infectives, say $I_{01},$ that appeared at $t=0$ in node 1 only:%
\begin{equation}
\begin{array}{cclcccl}
I_{1}(0) & = & I_{01}, &  & I_{2}(0) & = & 0, \\
S_{1}(0) & = & N_{1}-S_{12}(0)-I_{01}, &  & S_{2}(0) & = & N_{2}-S_{21}(0),
\\
I_{12}(0) & = & 0, &  & I_{21}(0) & = & 0, \\
S_{12}(0) & = & \frac{\gamma _{12}^{S}}{\gamma _{12}^{S}+\delta _{12}^{S}}%
N_{1} &  & S_{21}(0) & = & \frac{\gamma _{21}^{S}}{\gamma _{21}^{S}+\delta
_{21}^{S}}N_{2}.%
\end{array}
\label{ODE:ini}
\end{equation}

The choice for values for $S_{12}(0)$ and $S_{21}(0)$ will be explained
below in Section 3 by considering the migration processes before the
epidemic outbreak starts.

\section{Pure migration}

\label{sec:migration}

Consider migration of susceptibles before the epidemic starts in the
network. Setting $I_{1},I_{12},I_{2},I_{21}=0$ we obtain two decoupled
systems of ODEs for $S_{1},S_{12}$ and for $S_{2},S_{21}$ describing the
pure migration processes in the absence of an outbreak. Say, for the pair $%
S_{1},S_{12}$ we have
\begin{eqnarray}
\dot{S}_{1} &=&-\gamma _{12}^{S}S_{1}+\delta _{12}^{S}S_{12}
\label{eq:S1:migr} \\
\dot{S}_{12} &=&\hphantom{-}\gamma _{12}^{S}S_{1}-\delta _{12}^{S}S_{12}.
\label{eq:S12:migr}
\end{eqnarray}

Let migration start at $t=0$ with the initial conditions $S_{1}(0)=N_{1}$, $%
S_{12}(0)=0$. The solution to such an initial value problem is
\begin{equation}
S_{12}=N_{1}g_{12}^{S}(t),\qquad \ S_{1}=N_{1}-S_{12}
\label{eq:migr:solution}
\end{equation}%
where
\begin{equation}
g_{12}^{S}(t)=\frac{\gamma _{12}^{S}}{\gamma _{12}^{S}+\delta _{12}^{S}}%
\left[ 1-e^{-\left( \gamma _{12}^{S}+\delta _{12}^{S}\right) t}\right]
,\qquad \ t\geq 0  \label{eq:response}
\end{equation}%
is the response function (see below). Similar formulas are valid for the
second pair: $S_{2},S_{21}$. Thus, the number of migrants exponentially
tends to some limiting values
\begin{equation}
\lim_{t\rightarrow \infty }S_{12}=\frac{\gamma _{12}^{S}}{\gamma
_{12}^{S}+\delta _{12}^{S}}N_{1},\qquad \lim_{t\rightarrow \infty }S_{21}=%
\frac{\gamma _{21}^{S}}{\gamma _{21}^{S}+\delta _{21}^{S}}N_{2}.
\label{S12:lim}
\end{equation}%
These limits represent the dynamic equilibrium of migration processes in the
absence of the outbreak. At the equilibrium, the forward and backward
migrations fluxes compensate each other: $\dot{S}_{1\rightarrow 2}=\dot{S}%
_{1\leftarrow 2}$. So, in virtue of (\ref{eq:mig:rates}), $\gamma
_{12}^{S}S_{1}=\delta _{12}^{S}S_{12}$. Substituting $S_{1}=N_{1}-S_{12}$
and resolving with respect to $S_{12}$ yields (\ref{S12:lim}). We take
the equilibrium values from (\ref{S12:lim}) as the initial conditions for
the outbreak problem, that is reflected in (\ref{ODE:ini}).

The total populations in any node $S_{1}^{\Sigma }=S_{1}+S_{21}$ and $%
S_{2}^{\Sigma }=S_{2}+S_{12}$ are described as%
\begin{eqnarray*}
S_{1}^{\Sigma }(t) &=&N_{1}-N_{1}g_{12}(t)+N_{2}g_{21}(t) \\
S_{2}^{\Sigma }(t) &=&N_{2}-N_{2}g_{21}(t)+N_{1}g_{12}(t).
\end{eqnarray*}%
They can be non-monotonic for some choice of parameters. Next, $%
S_{1,2}^{\Sigma }$ asymptotically converges to
\begin{eqnarray*}
S_{1}^{\Sigma }(+\infty ) &=&N_{1}-N_{1}\frac{\gamma _{12}^{S}}{\gamma
_{12}^{S}+\delta _{12}^{S}}+N_{2}\frac{\gamma _{21}^{S}}{\gamma
_{21}^{S}+\delta _{21}^{S}} \\
S_{2}^{\Sigma }(+\infty ) &=&N_{2}-N_{2}\frac{\gamma _{21}^{S}}{\gamma
_{21}^{S}+\delta _{21}^{S}}+N_{1}\frac{\gamma _{12}^{S}}{\gamma
_{12}^{S}+\delta _{12}^{S}}.
\end{eqnarray*}%
If both centra are identical then their total population remains constant.

The migration dynamics described by this model seems reasonable. The
migration process resembles a diffusion process in physics, in which the
concentration tends monotonically to an equilibrium.

Note that if $\gamma _{12}^{S}\ll \delta _{12}^{S}$ then $S_{12}(t)\ll
N_{1},\forall t$, i.e.\ only a small share of the population from of node~1
is currently in node~2 (and vice verse: if $\gamma _{21}^{S}\ll \delta
_{21}^{S}$ then $S_{21}(t)\ll N_{2},\forall t$), which is appropriate for
large population centra. So, in this approximation $S_{mn}\lesssim (\gamma
_{mn}^{S}/\delta _{mn}^{S})S_{m}$.

To understand why function (\ref{eq:response}) can be associated with a
response function, consider a model for which the number of susceptibles can
vary even in the absence of migration due to other reasons (e.g., birth and
death). Let $\dot{N}_{1}$ be the rate of incoming ($\dot{N}_{1}>0$) or
outgoing ($\dot{N}_{1}<0$) individuals, i.e. the external source in the equations
\begin{equation}
\begin{array}{rcl}
\dot{S}_{1} & = & -\gamma _{12}^{S}S_{1}+\delta _{12}^{S}S_{12}+\dot{N}_{1}
\\
\dot{S}_{12} & = & \hphantom{-}\gamma _{12}^{S}S_{1}-\delta _{12}^{S}S_{12}%
\end{array}
\label{eq:N:variable}
\end{equation}%
with initial conditions $S_{1}(0)=N_{01}$, $S_{12}(0)=0$. Integrating the second equation
in view of relation $S_{1}=N_{1}-S_{12}$ yields:
\[
S_{12}(t)=\int_{0}^{t}\gamma _{12}^{S}\exp \left\{ -(\gamma _{12}^{S}+\delta
_{12}^{S})(t-t^{\prime })\right\} N_{1}(t^{\prime })\, \mathrm{d}t^{\prime
}\equiv \dot{g}_{12}^{S}(t)\ast N_{1}(t)
\]
where the asterisk denotes the convolution, $\dot{g}_{12}^{S}(t)$ is the
derivative of function (\ref{eq:response}): recall that $\dot{g}%
_{12}^{S}\ast H(t)=g_{12}^{S}$ where $H(t)$ is the unit-step Heaviside
function. So, $S_{12}=\dot{g}_{12}^{S}(t)\ast N_{1}(t)$ is the response in
the number of guests in node~2 on the population variation in node~1.

\section{Comparison with earlier models}
\label{sec:old}

The most epidemic network models deal with the
total number of infectives: $I_{n}^{\Sigma }=I_{n}+I_{mn}$, and susceptibles
$S_{n}^{\Sigma }=S_{n}+S_{mn}$ in node $n$. To compare these models take the
sum of Eqs.~(\ref{ODE:Sn}) and (\ref{ODE:In}) and obtain the equations
\[
\begin{array}{r@{}c@{}l}
\!\dot{S}_{n}^{\Sigma } &{}={}&-\beta _{n}S_{n}^{\Sigma }I_{n}^{\Sigma }%
-\gamma _{nm}^{S}S_{n}^{\Sigma
}+\gamma _{mn}^{S}S_{m}^{\Sigma }-(\delta _{mn}^{S}{-}\gamma
_{mn}^{S})S_{mn}+\left( \delta _{nm}^{S}{-}\gamma _{nm}^{S}\right) S_{nm}   \\
\!\dot{I}_{n}^{\Sigma } &{}={}&
\beta _{n}S_{n}^{\Sigma }I_{n}^{\Sigma
}-\alpha _{n}I_{n}^{\Sigma }-\gamma _{nm}^{I}I_{n}^{\Sigma }+\gamma
_{mn}^{I}I_{m}^{\Sigma }-\left( \delta _{mn}^{I}{-}\gamma _{mn}^{I}\right)\! I_{mn}{+}\left( \delta _{nm}^{I}{-}\gamma _{nm}^{I}\right)\!
I_{nm}.
\end{array}
\]%
Thus, we cannot obtain equations for total numbers of species only: they become coupled with the the equations for guest individuals.
Even when the number of guests is relatively small $S_{mn}\ll S_{n}$, $%
I_{mn}\ll I_{n}$, and the approximation $I_{n}^{\Sigma }\approx I_{n}$, $%
S_{n}^{\Sigma }\approx S_{n}$ holds, we cannot neglect terms with $\delta $'s
because $S_{mn}\ll S_{n}$, $I_{mn}\ll I_{n}$ are not always can be valid.
In fact, the terms with $\gamma $'s and $\delta $'s may be of the same order.
This complicates the model but makes it more realistic.

Many authors simply insert terms proportional to the relevant population
size in neighbouring nodes to provide coupling between SIR centra:
\begin{eqnarray}
\dot{S}_{n} &=&-\beta _{n}S_{n}I_{n}\hphantom{{}-\alpha _{n}I_{n}}\quad
{}+\chi _{mn}^{S}S_{m}  \label{ODE:Sn:standard} \\
\dot{I}_{n} &=&\hphantom{-}\beta _{n}S_{n}I_{n}-\alpha _{n}I_{n}\quad
{}+\chi _{mn}^{I}I_{m}  \label{ODE:In:standard}
\end{eqnarray}%
where $\chi _{mn}^{S,I}\geq 0$ are coupling coefficients (cf.\ \cite{DG,Wang04}).
In the case of pure migration between two centra ($\alpha _{n}=\beta _{n}=0$%
, $I_{n}=0$) they are reduced to%
\begin{equation}
\dot{S}_{1}=\chi _{12}^{S}S_{2},\qquad \dot{S}_{2}=\chi _{21}^{S}S_{1}.
\label{eq:pure:migr:standard}
\end{equation}%
Note that this model does not guarantee preservation of the total population
size because the sum $S_{1}+S_{2}$ is variable
\[
\dot{S}_{1}+\dot{S}_{2}=\chi _{12}^{S}S_{2}+\chi _{21}^{S}S_{1},
\]%
which is unrealistic. By eliminating one variable we see that the system has
unstable dynamics%
\[
\ddot{S}_{1}=\chi _{12}^{S}\chi _{21}^{S}S_{1}\Longrightarrow S_{1}=Ae^{\chi
t}+Be^{-\chi t},\qquad \chi =\sqrt{\chi _{12}^{S}\chi _{21}^{S}}>0,
\]%
i.e., a growing particular solution. Thus, the traditional approach does not
describe the migration between centra properly. Although this instability
can potentially be hidden in the background of the outbreak and not be
observable in certain epidemic model scenarios.

Nevertheless the model can be easily corrected by introducing inverse fluxes  (cf.\ \cite{Wang03,Arino05})%
\begin{eqnarray}
\dot{S}_{n} &=&-\beta _{n}S_{n}I_{n}\hphantom{{}-\alpha _{n}I_{n}}\quad
{}+\chi _{mn}^{S}S_{m}-\chi _{nm}^{S}S_{n}  \label{OSE:Sn:corrected} \\
\dot{I}_{n} &=&\hphantom{-}\beta _{n}S_{n}I_{n}-\alpha _{n}I_{n}\quad
{}+\chi _{mn}^{I}I_{m}-\chi _{nm}^{I}I_{n}.  \label{OSE:In:corrected}
\end{eqnarray}

Then for a pure migration model we have%
\begin{equation}
\dot{S}_{1}=-\chi _{21}^SS_{1}+\chi _{12}^{S}S_{2},\qquad \dot{S}_{2}=\chi
_{21}^{S}S_{1}-\chi _{12}^{S}S_{2}  \label{ODE:corrected:pure:migr}
\end{equation}%
implying $S_{1}+S_{2}=const$. The solutions of ODEs (\ref%
{ODE:corrected:pure:migr}) with initial conditions $S_{1}(0)=N_{1}$, $%
S_{2}(0)=N_{2}$ demonstrate exponential, diffusion-like behaviour of each
node population%
\begin{eqnarray*}
S_{1} &=&\frac{\chi _{mn}^{S}(N_{1}+N_{2})}{\chi _{mn}^{S}+\chi _{nm}^{S}}+%
\frac{\chi _{nm}^{S}N_{1}-\chi _{mn}^{S}N_{2}}{\chi _{mn}^{S}+\chi _{nm}^{S}}%
\exp \left[ -\left( \chi _{mn}^{S}+\chi _{nm}^{S}\right) t\right] \\
S_{2} &=&\frac{\chi _{nm}^{S}(N_{1}+N_{2})}{\chi _{mn}^{S}+\chi _{nm}^{S}}+%
\frac{\chi _{mn}^{S}N_{2}-\chi _{nm}^{S}N_{1}}{\chi _{mn}^{S}+\chi _{nm}^{S}}%
\exp \left[ -\left( \chi _{mn}^{S}+\chi _{nm}^{S}\right) t\right].
\end{eqnarray*}%
In the case $\chi _{mn}^{S}=\chi _{nm}^{S}$ both solutions tend to $%
S_{1}(+\infty )=S_{2}(+\infty )=\frac{1}{2}\left( N_{1}+N_{2}\right) $, i.e.
their populations become equal (fully mixed). Thus the dynamics of the
corrected model seems to be more realistic but nevertheless does not satisfy an
intuitive interpretation of the equilibrium of the migration process.

In \cite{SKG11a}, in order to obtain more realistic migration dynamics,
different flux terms are added to equations (\ref{ODE:Sn})--(\ref{ODE:In})
in the form of convolutions (transition terms)
\[
S_{n\rightarrow m}=\dot{g}_{nm}^{S}\ast S_{n},\qquad I_{n\rightarrow m}=\dot{%
g}_{nm}^{I}\ast I_{n}
\]%
where $g$'s are the relevant response functions. Then, the dynamics is
described by integro-differential equations%
\begin{eqnarray}
\dot{S}_{n} &=&-\beta _{n}S_{n}I_{n}\hphantom{{}-\alpha _{n}I_{n}}\quad {}-%
\textstyle\frac{d}{dt}(\dot{g}_{nm}^{S}\ast S_{n})+\frac{d}{dt}(\dot{g}%
_{mn}^{S}\ast S_{m})  \label{eq:Sn:response} \\
\dot{I}_{n} &=&\hphantom{-}\beta _{n}S_{n}I_{n}-\alpha _{n}I_{n}\quad {}-%
\textstyle\frac{d}{dt}(\dot{g}_{nm}^{I}\ast I_{n})+\frac{d}{dt}(\dot{g}%
_{mn}^{I}\ast I_{m}).  \label{eq:In:response}
\end{eqnarray}

A natural choice for the response functions is the exponential form
\begin{equation}
g_{mn}^{I,S}(t)=\varepsilon _{mn}^{I,S}\left[ 1-e^{-t/\tau _{mn}^{I,S}}%
\right]  \label{ken2}
\end{equation}%
where $\tau $'s are the characteristic migration times, $\varepsilon $'s are
coupling parameters, $t\geq 0$. The form of these response functions is the
same as in (\ref{eq:response}) obtained solving the initial value problem,
see Section~\ref{sec:migration}

Note that for a response function in the form of (\ref{ken2}), the
integro-differential equations (\ref{eq:Sn:response})--(\ref{eq:In:response}%
) can be reduced to ODEs. We introduce additional variables $S_{nm}=\dot{g}%
_{nm}^{S}\ast S_{n}$, $I_{nm}=\dot{g}_{nm}^{I}\ast I_{n}$ which aim to
capture the number of guests in node $m$ coming from node $n$, in agreement
with notations used in the present work. They obey the following ODEs
\begin{equation}
\dot{S}_{nm}+\frac{1}{\tau _{nm}^{S}}S_{nm}=\frac{\varepsilon _{nm}^{S}}{%
\tau _{nm}^{S}}S_{n},\qquad \dot{I}_{nm}+\frac{1}{\tau _{nm}^{I}}I_{nm}=%
\frac{\varepsilon _{nm}^{I}}{\tau _{nm}^{I}}I_{n}  \label{eq:ODE:F}
\end{equation}
that can be easily checked. Then ODEs
\begin{eqnarray}
\dot{S}_{n} &=&-\beta _{n}S_{n}I_{n}\hphantom{{}-\alpha _{n}I_{n}}\quad {}-%
\dot{S}_{nm}+\dot{S}_{mn} \\
\dot{I}_{n} &=&\hphantom{-}\beta _{n}S_{n}I_{n}-\alpha _{n}I_{n}\quad {}-%
\dot{I}_{nm}+\dot{I}_{mn}
\end{eqnarray}%
together with ODEs (\ref{eq:ODE:F}) form a closed system of equations for $%
n=1,2$ and $m=2,1$.

For a pure migration model (neglecting the outbreak dynamics) we obtain the
following ODEs%
\begin{equation}
\begin{array}{ccccccc}
\dot{S}_{1} & = & -\dot{S}_{12}+\dot{S}_{21}, &  & \dot{S}_{12}+S_{12}/\tau
_{12}^{S} & = & \varepsilon _{12}^{S}S_{1} \\
\dot{S}_{2} & = & \dot{S}_{12}-\dot{S}_{21}, &  & \dot{S}_{21}+S_{21}/\tau
_{21}^{S} & = & \varepsilon _{21}^{S}S_{2}%
\end{array}
\label{eq:old:migration}
\end{equation}%
with initial conditions $%
S_{1}(0)=N_{1},S_{2}(0)=N_{2},S_{12}(0)=S_{21}(0)=0. $

Taking the sum of the two left equations we see that the total population is
preserved: $S_{1}+S_{2}=const=N_{1}+N_{2}$.

In the case of identical migration parameters for the both nodes: $\tau
_{12}^{S}=\tau _{21}^{S}=\tau $, $\varepsilon _{12}^{S}=\varepsilon
_{21}^{S}=\varepsilon $ the number of migrants in node 2 is described as in (%
\ref{eq:migr:solution})--(\ref{eq:response}):
\[
S_{12}=N_{1}\frac{\varepsilon }{1+2\varepsilon }\left( 1-\exp \left\{ -\frac{%
1+2\varepsilon }{\tau }t\right\} \right).
\]%
In the general case the solution can be represented via two exponential
functions with different characteristic times $\tau _{1}$ and $\tau _{2}$
and not necessarily monotonic (which is quite unrealistic). Nevertheless,
neglecting backward migration from node 2 to node 1 by setting $S_{21}=0$,
i.e. solving equations $\dot{S}_{1}=-\dot{S}_{12}$, $\dot{S}%
_{12}+S_{12}/\tau _{12}^{S}=\varepsilon _{12}^{S}S_{1}$, yields exactly the
solution (\ref{eq:migr:solution})--(\ref{eq:response}):%
\[
S_{12}=N_{1}g_{12}^{S}=N_{1}\varepsilon _{12}^{S}\left[ 1-e^{-t/\tau
_{12}^{S}}\right] ,\ S_{1}=N_{1}-S_{12}.
\]

Summing up, we conclude that the model proposed in the present work behaves
appropriately: at first populations increase exponentially, then they tend
monotonically to their final values, say $S_{1}(+\infty )=N_{1}-\varepsilon
_{mn}^{S}N_{1}$. Other useful approximations will be applied in specific
situations.

Next we compare the response function (\ref{ken2}) defined in \cite%
{SKG08,SKG11a} and the response function (\ref{eq:response}), and express
the basic migration parameters such as the migration characteristic time $%
\tau _{12}^{S}$ and the coupling coefficient $\varepsilon _{12}^{S}$ via the
migration parameters $\gamma _{12}^{S}$ and $\delta _{12}^{S}$ introduced
here:
\begin{equation}
\tau _{12}^{S}=\frac{1}{\gamma _{12}^{S}+\delta _{12}^{S}},\qquad
\varepsilon _{12}^{S}=\frac{\gamma _{12}^{S}}{\gamma _{12}^{S}+\delta
_{12}^{S}}  \label{eq:tau-eps}
\end{equation}%
The inverse relations are
\[
\gamma _{12}^{S}=\frac{\varepsilon _{12}^{S}}{\tau _{12}^{S}},\qquad \delta
_{12}^{S}=\frac{1-\varepsilon _{12}^{S}}{\tau _{12}^{S}}.
\]

Coefficient $\varepsilon _{12}^{S}$ represents a share of the population
from node 1 migrated to node 2 at dynamical equilibrium or a share of time
the individuals from node 1 spend in node 2 on average. In the case of small
coupling, the estimation of the order of different terms is very useful.

Analogous response functions can be defined for all other population
classes, they determine the dynamics due to pure migration when the disease
transmission and removal is disregarded by setting $\alpha $'s and $\beta $%
's to zero. The response function for the migration of susceptibles and
infectives
\begin{equation}
g_{12}^{S,I}(t)=\varepsilon _{12}^{S,I}\left[ 1-e^{-t/\tau _{12}^{S,I}}%
\right] ,\quad \tau _{12}^{S,I}=\frac{1}{\gamma _{12}^{S,I}+\delta
_{12}^{S,I}},\quad \varepsilon _{12}^{S,I}=\frac{\gamma _{12}^{S,I}}{\gamma
_{12}^{S,I}+\delta _{12}^{S,I}}  \label{ken2a}
\end{equation}%
will be used intensively below.

\begin{figure}[h]
\begin{center}
\includegraphics[width=120mm]{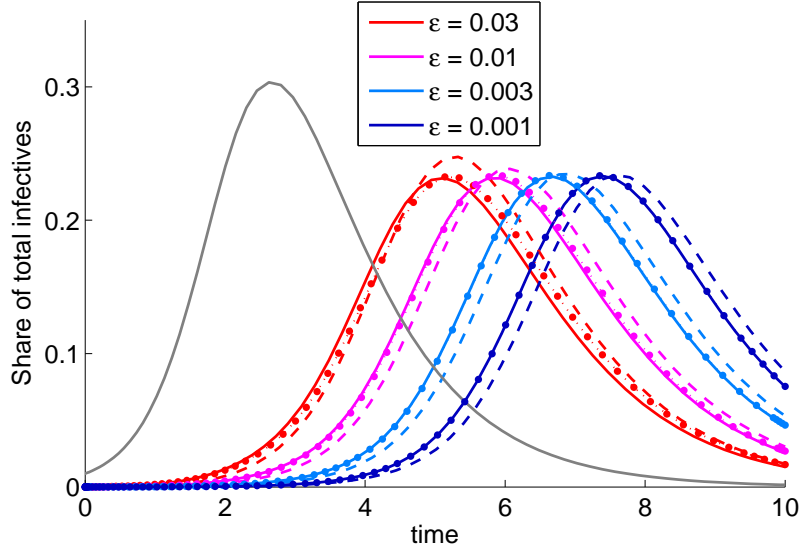}
\end{center}
\caption{Dynamics of the total number of infectives $I^{\Sigma }_{2}\equiv
I_2+I_{12}$ in the second node (divided by its population): (a) computed via
the full equations (\protect\ref{ODE:Sn})--(\protect\ref{ODE:Inm}) (colored
solid line); by the SIC approximation (\protect\ref{ode:I2})--(\protect\ref%
{ode:S21}) (dotted lines); assuming the absence of guest susceptibles before the
outbreak (dashed lines). The curves are plotted for different values of the
coupling coefficient $\protect\varepsilon _{12}^{I} \equiv \protect\gamma %
_{12}^{I}/(\protect\gamma _{12}^{I}+\protect\delta _{12}^{I})$ (indicated
in the legend). Parameters: $N_{2}/N_{1}=0.8$, $\protect\alpha _{1}=%
\protect\alpha _{2}=1$, $\protect\beta _{1}=3,\protect\beta _{2}=2.5$, $%
\protect\tau _{12}^{I,S} \equiv 1/(\protect\gamma _{12}^{I,S}+\protect\delta %
_{12}^{I,S})=3$, $\protect\varepsilon _{12}^{S}=\protect\varepsilon %
_{21}^{I,S}=\protect\varepsilon _{12}^{I}$. The epidemic outbreak in the
first node $I^{\Sigma }_{1}\equiv I_1+I_{21}$ (divided by its population) is
indicated by a grey line.}
\label{fig:SIR2}
\end{figure}

Examples of numerical solutions to the initial value problems (\ref{ODE:Sn}%
)--(\ref{ODE:Inm})--(\ref{ODE:ini}) are shown in Figure~\ref{fig:SIR2}.

\section{Small initial contagion (SIC) approximation}

\label{sec:sic}

In many situations the number of external infectives triggering an epidemic
outbreak in a given center is small compared to the number of infectives
occurring during the developed outbreak. If this is the case, the model can
be simplified in the framework of the small initial contagion (SIC)
approximation introduced in \cite{SKG08,SKG11a}.

In the SIC approximation we can split the outbreak process in every node
into two stages: (i) contamination and (ii) the developed outbreak. At the
contamination stage, the number of infectives is relatively small whereas
the total population consists mainly of susceptibles. Then on the r.h.s. of (%
\ref{ODE:Sn})--(\ref{ODE:Inm}), $S_{1}\approx N_1$ and $S_{2}\approx N_2$,
the equations become linear and can be easily analyzed. At
the second stage every node become non-sensitive to small migration
processes and the dynamics can be well described by a standard SIR process.

In the SIC approximation, the initial number of infectives should be
small that obviously can occur in a large
population center. Thus the first conditions should be
\begin{equation}
N_{n}\gg 1,\qquad \forall n.  \label{sic:N>>1}
\end{equation}

Secondly, the coupling between nodes should be small, so that in the
developed outbreak the flux of infective migrants remains small compared to
the population of nodes to which those infectives travel. This implies that
all $\varepsilon $'s (see (\ref{ken2a})) are small:%
\begin{equation}
\varepsilon _{12}^{S}=\frac{\gamma _{12}^{S}}{\gamma _{12}^{S}+\delta
_{12}^{S}}\ll 1\Leftrightarrow \gamma _{mn}^{I,S}\ll \delta
_{mn}^{I,S},\qquad \forall m\neq n.  \label{sic:eps<<1}
\end{equation}

Next, consider a network of population centers. Let an initial number of
infectives $I_{0n}$ suddenly appear in one of the nodes, say in node $n$.
Then $I_{0n}\ll I_{bn}$ should hold where $I_{bn}=\max \left\{
I_{n}(t)\right\} $ is the number of infectives in the peak of the outbreak. But if
infectives are arriving gradually into a node (which is almost always the
case for many nodes in the network), then it is not their total number that
is essential but some effective number of initial infectives $I_{0n}^{%
\mathrm{eff}}$: earlier immigrated infectives have time to contaminate more
local susceptibles than ones immigrated later. So, the effective number of
initial infectives $I_{0n}^{\mathrm{eff}}$ should be a weighted integral of $%
I_{nm}(t)$ (see below): $I_{0n}^{\mathrm{eff}}\ll I_{bn}$.

Note if the basic reproduction number defined as
\begin{equation}
\rho _{n}=\frac{\beta _{n}N_{n}}{\alpha _{n}}  \label{rho-n}
\end{equation}%
is not close to unity then $I_{bn}\sim N_{n}$. Thus, the additional and the
most important condition to maintain the validity of the SIC approximation is%
\begin{equation}
I_{0n}^{\mathrm{eff}}\ll N_{n}  \label{sic:Ieff<<N}
\end{equation}%
where $I_{0n}^{\mathrm{eff}}$ is defined below by (\ref{eq:Ieff}) for the
two-nodes network or by (\ref{eq:Ieff-network}) for a general network.

If the reproduction number $\rho _{n}$ only slightly exceeds unity, the
number of infectives in the outbreak is estimated as follows (cf.\ \cite{DG}%
)
\begin{equation}
I_{bn}\approx -\frac{N_{n}\ln (1-I_{0n}/N_{n})}{\rho _n}+\frac{N_{n}}{2}\left(
\rho _{n}-1\right) ^{2}+O\left[ \left( \left( \rho _{n}-1\right) ^{3}\right) %
\right].   \label{Ibn:small:rho}
\end{equation}%
So the relation $I_{bn}\sim O(N_{n})$ holds if (a) $I_{0n}\sim N_{n}$ (in
this case the outbreak is not evident as $I_{bn}-I_{0n}\ll I_{0n}$) and (b)
if
\begin{equation}
\rho _{n}-1\gg N_{n}^{-1/2}.  \label{sic:r-1>>1/sqrt(N)}
\end{equation}%
We conclude that the SIC approximation is reasonable for epidemic models
under above condition (\ref{sic:r-1>>1/sqrt(N)}).

In the SIC approximation, i.e.\ when conditions (\ref{sic:N>>1}), (\ref%
{sic:eps<<1}), (\ref{sic:Ieff<<N}), (\ref{sic:r-1>>1/sqrt(N)}) hold,
migration fluxes are small. They cannot change dramatically the populations
at all nodes. However, the fluxes of infectives from a node with an outbreak
to a non-contaminated node are essential (as these fluxes trigger the
outbreak in that node or another node at an early stage of outbreak
development).

Consider the initial value problem (\ref{ODE:Sn})--(\ref{ODE:Inm})--(\ref%
{ODE:ini}) in the SIC approximation.
As node 1 is contaminated first, it is not sensitive to the outbreak in node~2 which will develop after a certain delay.

To build asymptotic we assume that all coupling coefficients $\varepsilon
_{mn}^{S,I}$ are of the same order with respect to a small parameter $%
\varepsilon $: $\varepsilon _{mn}^{S,I}\equiv \gamma _{mn}^{S,I}/(\gamma
_{mn}^{S,I}+\delta _{mn}^{S,I})$ $=O(\varepsilon )$. Also we assume that $%
I_{mn}=O(\varepsilon )I_{m}$, $S_{mn}=O(\varepsilon )S_{m}$, $\delta
_{mn}^{S,I}=O(1)\alpha _{m,n}=O(1)\beta _{m,n}N_{m,n}$. Also at
contamination stage $I_{2}<O(1)I_{21,12}$, $S_{2}\approx N_{2}-S_{21}$.

We rewrite equation (\ref{ODE:Sn})--(\ref{ODE:In}) for host species in node~1 separating $O(\varepsilon )$ terms and enclosing them  in curly brackets
\begin{eqnarray*}
\dot{S}_{1} &=&-\beta _{1}S_{1}I_{1}\hphantom{-(\alpha_1))}\quad {}+\left\{
-\beta _{1}S_{1}I_{21}-\gamma _{12}^{S}S_{1}+\delta _{12}^{S}S_{12}\right\}
\\
\dot{I}_{1} &=&\hphantom{-}\left( \beta _{1}S_{1}-\alpha _{1}\right)
I_{1}\quad {}+\left\{ \beta _{1}S_{1}I_{21}-\gamma _{12}^{I}I_{1}+\delta
_{12}^{I}I_{12}\right\}.
\end{eqnarray*}%
Neglecting terms in curly brackets
we see that outbreak in node~1 can be described by standard SIR model for an
isolated node:%
\begin{eqnarray}
\dot{S}_{1} &=&-\beta _{1}S_{1}I_{1}  \label{sic:S1} \\
\dot{I}_{1} &=&\hphantom{-}\beta _{1}S_{1}I_{1}-\alpha _{1}I_{1}.
\label{sic:I1}
\end{eqnarray}

Now we rewrite equation (\ref{ODE:Sn})--(\ref{ODE:In}) for host species in
node~2
\begin{eqnarray}
\dot{S}_{2} &=&-\beta _{2}S_{2}(I_{2}+I_{12})
+\left\{ -\gamma _{21}^{S}S_{2}+\delta _{21}^{S}S_{21}\right\} \label{sic:S2}\\
\dot{I}_{2} &=&\hphantom{-}\left( \beta _{2}S_{2}-\alpha _{2}\right) I_{2}+%
\left[ \beta _{2}S_{2}I_{12}\right] \quad {}+\left[ \delta _{21}^{I}I_{21}%
\right] +\left\{ -\gamma _{21}^{I}I_{2}\right\}. \label{sic:I2}
\end{eqnarray}%
Here the term remaining always small and to be neglected is enclosed in
curly brackets. Small terms which can prevail at the stage of contamination
when $I_{2}$ is small are enclosed in square brackets. They are coupling
terms and represent two fluxes: $\mu _{1}(t)=\beta _{2}S_{2}I_{12}$ and $\mu
_{2}(t)=\delta _{21}^{I}I_{21}$. Flux $\mu _{1}$ is due to the infected
individuals belonging to node 1 and currently migrated to node 2
contaminating susceptibles there. Flux $\mu _{2}$ is due to susceptible
individuals migrated to node 1 from node 2, contaminated their and returning
as infectives their host node.

Rewrite equation (\ref{ODE:Inm}) for $I_{12}$:
\begin{equation}\label{sic:I12}
\dot{I}_{12}=-\left( \alpha _{2}+\delta _{12}^{I}\right) I_{12}+\gamma
_{12}^{I}I_{1}\hphantom{-}\quad {}+\left\{ \beta
_{2}S_{12}(I_{2}+I_{12})\right\}.
\end{equation}%
Terms in curly brackets are small and can be neglected.

Now we rewrite analogous equation for $I_{21}$%
\begin{equation} \label{sic:I21}
\dot{I}_{21}=-\left( \alpha _{1}+\delta _{21}^{I}\right) I_{21}+ \left[\beta
_{1}S_{21}I_{1}\right] \quad {}+\left\{ \gamma _{21}^{I}I_{2}+\beta
_{1}S_{21}I_{21}\right\}.
\end{equation}%
We can neglect terms in curly brackets but essential term in the square brackets should remain.

The value of $S_{21}(t)$ initially equals $\varepsilon
_{21}^{S}=\gamma _{21}^{S}/\left( \gamma _{21}^{S}+\delta _{21}^{S}\right) $
but can vary during contamination stage. Its varying is described by equation
(\ref{ODE:Snm})%
\begin{equation}\label{sic:S21}
\dot{S}_{21}=\left( -\beta _{1}I_{1}-\delta _{21}^{S}\right) S_{21}%
\quad {}+\gamma _{21}^{S}S_{2}+\left\{
-\beta _{1}S_{21}I_{21}\right\}
\end{equation}
It can vary noticeably during contamination stage for node 2.

At the initial stage we can approximate $S_{2}\approx
N_{2}-S_{21}$%
\[
\dot{S}_{21}+\left( \beta _{1}I_{1}(t)+\delta _{21}^{S}\right) S_{21}=\gamma
_{21}^{S}\left( N_{2}-S_{21}\right)
\]%
or%
\begin{equation}\label{eq:S21}
\dot{S}_{21}+\left( \beta _{1}I_{1}(t)+\left( \delta _{21}^{S}+\gamma
_{21}^{S}\right) \right) S_{21}=\gamma _{21}^{S}N_{2}.
\end{equation}%
Re-writing (\ref{sic:S1}) in the form $\beta _{1}I_{1}=-\dot{S}_{1}/S_{1}$ and
also utilizing (\ref{eq:tau-eps}) we can write (\ref{eq:S21}) as
\[
\dot{S}_{21}+\left( -\frac{\dot{S}_{1}}{S_{1}}+\frac{1}{\tau _{21}^{S}}%
\right) \dot{S}_{21}=\frac{S_{21}(0)}{\tau _{21}^{S}}
\]%
where $S_{21}(0)=\varepsilon _{21}^{S}N_{2}$. When $S_{1}=const$ it has a
steady-state solution $S_{21}=S_{21}(0)$. In the case of the outbreak the
solution can be written in the quadrature from%
\[
S_{21}(t)=S_{21}(0)\left( 1+\frac{1}{\tau _{21}^{S}}\int_{0}^{t}e^{t^{\prime
}/\tau _{21}^{S}}\frac{N_{1}dt^{\prime }}{S_{1}(t^{\prime })}\right) \frac{%
S_{1}(t)}{N_{1}}e^{-t/\tau _{21}^{S}}.
\]

Coupling with node 1 is essential in the SIC approximation at the
initial stage only before the developed outbreak. At this stage we solve (\ref%
{sic:I2})--(\ref{sic:S21}) neglecting terms in the curly brackets (then it becomes a linear inhomogeneous system of ODEs)
approximating $S_{2}\approx N_{2}$:
\begin{eqnarray}
\dot{I}_{2}-\left( \beta _{2}N_{2}-\alpha _{2}\right) I_{2} &=&\beta
_{2}N_{2}I_{12}+\delta _{21}^{I}I_{21}  \label{ode:I2} \\
\dot{I}_{12}+(\alpha _{2}+\delta _{12}^{I})I_{12} &=&\gamma _{12}^{I}I_{1}
\label{ode:I12} \\
\dot{I}_{21}+\left( \alpha _{1}+\delta _{21}^{I}-\beta
_{1}N_{2}S_{21}\right) I_{21} &=&\beta _{1}N_{2}\left( S_{21}I_{1}\right)
\label{ode:I21} \\
\dot{S}_{21}+\left( \beta _{1}I_{1}(t)+\left( \delta _{21}^{S}+\gamma
_{21}^{S}\right) \right) S_{21} &=&\gamma _{21}^{S}N_{2}.  \label{ode:S21}
\end{eqnarray}

We re-write first three equations in the simplest form introducing new
parameters%
\begin{eqnarray}
\dot{I}_{2}-\lambda _{2}I_{2} &=&a_{12}I_{12}+a_{21}I_{21} \label{eq1}\\
\dot{I}_{12}+\lambda _{12}I_{12} &=&b_{12}I_{1} \label{eq2}\\
\dot{I}_{21}+\lambda _{21}I_{21} &=&b_{21}\left( S_{21}I_{1}\right) \label{eq3}
\end{eqnarray}%
where $\lambda _{2}=\beta _{2}N_{2}-\alpha _{2}=\left( \rho _{2}-1\right)
\alpha _{2}$ is initial growth rate in node 2; $\lambda _{12}=(\alpha
_{2}+\delta _{12}^{I})$ and $\lambda _{21}=\left( \alpha _{1}+\delta
_{21}^{I}-\beta _{1}\bar{\varepsilon}_{21}^{S}N_{2}\right) $\ are initial
decay rate of guest species $I_{21}$ and $I_{21}$, respectively; $%
a_{12}=\beta _{2}N_{2}$, $a_{21}=\delta _{21}^{I}$, $b_{12}=\gamma _{12}^{I}$%
,\ $b_{21}=\beta _{1}N_{2}$. Note in the SIC approximation should be $\beta
_{1}\bar{\varepsilon}_{21}^{S}N_{2}\ll \alpha _{1},\delta _{21}^{I}$
therefore $\lambda _{21}\approx \alpha _{1}+\delta _{21}^{I}>0$.

Solving system (\ref{eq1})--(\ref{eq3}) by the Laplace transform
method we obtain%
\begin{eqnarray*}
I_{12} &=&b_{12}I_{1}\ast e^{-\lambda _{12}t},\qquad
I_{21} \quad=\quad b_{21}I_{1}\ast e^{-\lambda _{21}t} \\
I_{2} &=&C_{12}^{(1)}I_{1}\ast e^{\lambda _{2}t}+C_{21}^{(2)}\left( S_{21}I_{1}\right)
\ast e^{\lambda _{2}t}\\
&& \hspace{1cm}{}-\frac{a_{12}b_{12}}{\lambda _{2}+\lambda _{12}}%
I_{1}\ast e^{-\lambda _{12}t}-\frac{a_{21}b_{21}}{\lambda _{2}+\lambda _{21}}%
\left( S_{21}I_{1}\right) \ast e^{-\lambda _{21}t}
; \\
C_{12}^{(1)} &=&\frac{a_{12}b_{12}}{\lambda _{2}+\lambda _{12}},\qquad C_{21}^{(2)}\;{{}={}}\;\frac{%
a_{21}b_{21}}{\lambda _{2}+\lambda _{21}}
\end{eqnarray*}%
where $\ast $ denotes the convolution. At time $t\gtrsim \lambda _{2}^{-1}$
only the growing terms for $I_{2}$ are essential, and the simplified expression
takes the form
\begin{equation}
I_{2}\simeq C_{12}^{(1)}\int_{0}^{t}I_{1}(t^{\prime })e^{\lambda _{2}\left(
t-t^{\prime }\right) }\ \mathrm{d}t^{\prime
}+C_{21}^{(2)}\int_{0}^{t}S_{21}(t^{\prime })I_{1}(t^{\prime })e^{\lambda
_{2}\left( t-t^{\prime }\right) }\ \mathrm{d}t^{\prime }
\label{eq:I2}
\end{equation}%
where
\begin{equation}
C_{12}^{(1)}=\frac{\gamma _{12}^{I}\beta _{2}N_{2}}{\beta _{2}N_{2}+\delta
_{12}^{I}}, \qquad C_{21}^{(2)}=\frac{\delta _{21}^{I}\beta _{1}}{\beta _{2}N_{2}-\alpha
_{2}+\alpha _{1}+\delta _{21}^{I}}.  \label{eq:C2}
\end{equation}%
Constant $C_{12}^{(1)} = O(\varepsilon)$ as it is proportional to $\gamma _{12}^{I}$.
Constant $C_{21}^{(2)}$ is not small but the integrand contains value $S_{21}$ which is proportional to $\gamma _{21}^{S}$, i.e. also has order $O(\varepsilon)$.

%
%

As $I_{1}(t)$ in the integrand describes an outbreak in node 1 which is a
decaying function after reaching the maximal outbreak in the node, we see
that contribution of it in integral (\ref{eq:I2}) is negligible after time $%
t_{\ast }$: $t_{b1}<t_{\ast }$. Thus for $t>t_{\ast }$ function $I_{2}(t)$
grows exponentially as in the standard SIR model%
\begin{eqnarray}
I_{2} &\simeq &e^{\lambda _{2}t}\int_{0}^{t_{\ast }}\left(
C_{12}^{(1)}+C_{21}^{(2)}S_{21}(t^{\prime })\right) I_{1}(t^{\prime })e^{-\lambda
_{2}t^{\prime }}\, \mathrm{d}t^{\prime }  \nonumber \\
&\approx &e^{\lambda _{2}t}\int_{0}^{+\infty }\left(
C_{12}^{(1)}+C_{21}^{(2)}S_{21}(t^{\prime })\right) I_{1}(t^{\prime })e^{-\lambda
_{2}t^{\prime }}\, \mathrm{d}t^{\prime }=I_{02}^{\mathrm{eff}}e^{\lambda
_{2}t}.  \label{eq:I2:exp}
\end{eqnarray}%
Thus, we have the following evaluation for the effective number of
infectives in node 2:%
\begin{equation}
I_{02}^{\mathrm{eff}}\approx \int_{0}^{+\infty }
\left[ C_{12}^{(1)}I_{1}(t)+C_{21}^{(2)}S_{21}(t)I_{1}(t)\right] e^{-\lambda
_{2}t}\
\, \mathrm{d}t.  \label{eq:Ieff}
\end{equation}%
Calculation of integral (\ref{eq:Ieff}) needs simple approximation of
solutions for the standard SIR model. It will be considered in a separate
work.

\section{Outbreak time}

\label{sec:outbreak-time}

Next, we evaluate the outbreak time $t_{bn}$, the time from the introduction
of infection up until the peak of the outbreak, in every node in the
framework of the SIC approximation. Let the initial growth in the node 1 be
an exponential:


\begin{equation}
I_{1}\approx I_{01}e^{\lambda _{1}t}  \label{I1lin}
\end{equation}%
where $\lambda _{1}$ is the initial growth rate of infectives in node 1
\begin{equation}
\lambda _{1}=\beta _{1}S_{1}(0)-\alpha _{1}\approx \beta _{1}N_{1}-\alpha
_{1}=\alpha _{1}\left( \rho _{1}-1\right) .  \label{lambda1}
\end{equation}

Subsequent behaviour in the first center can be approximated by the limiting
solution $i^{\lim }\left( t;\rho \right) $ introduced and described in \cite%
{SKG08,SKG11a}
\begin{equation}
I_{1}=N_{1}i^{\lim }\left[ \alpha _{1}\left( t-t_{b1}(I_{01})\right) ;\rho
_{1}\right]  \label{I1lim}
\end{equation}%
where $i=I/N$ is the share of infectives in the node.

The outbreak time $t_{b1}(I_{01})$ can be roughly approximated by \cite%
{SKG08,SKG11a}
\begin{equation}
t_{b1}=\frac{1}{\lambda _{1}}\ln \frac{A_{0}N_{1}}{I_{01}}\approx \frac{1}{%
\lambda _{1}}\ln \frac{\lambda _{1}N_{1}}{\alpha _{1}I_{01}}.  \label{t1outb}
\end{equation}%
where $A_{0}(\rho _{1})$ is the parameter of the large negative time
asymptotics of the limiting solution:
\[
i^{\lim }\rightarrow A_{0}e^{\lambda _{1}\left( t-t_{b1}\right) },\
t\rightarrow -\infty .
\]%
This shows the share of infectives at the instant of contamination that is
required to trigger an outbreak under the assumption that the initial
exponential growth continues up to the moment of the peak of the outbreak.

The dynamics of $I_{2}$ is described by (\ref{eq:I2}) at the initial stages
before the developed outbreak: at the contamination stage when the migration
of infectives from node 1 is essential and at the stage of exponential
growth. At this stage (\ref{eq:I2:exp}) the coupling can be neglected and
the dynamics is described by a limiting solution

\begin{equation}
I_{2}=N_{2}i^{\lim }\left[ \alpha _{2}\left( t-t_{b2}\right) ;\rho _{2}%
\right]\approx N_{2}A_{0}e^{\lambda _{2}\left( t-t_{b2}\right) }.
\label{eq:I2:lim}
\end{equation}

Comparing (\ref{eq:I2:exp}) and (\ref{eq:I2:lim}) we elaborate that%
\[
t_{b2}=\frac{1}{\lambda _{2}}\ln \frac{A_{0}N_{2}}{I_{02}^{\mathrm{eff}}}%
\approx \frac{1}{\lambda _{2}}\ln \frac{\lambda _{2}N_{2}}{\alpha
_{2}I_{02}^{\mathrm{eff}}} .
\]

\section{Epidemic spread in a 1D lattice of coupled SIR nodes}

In the case of a general network of $M$ interacting nodes Eqs.~(\ref{ODE:Sn}%
)--(\ref{ODE:In})--(\ref{ODE:Snm})--(\ref{ODE:Inm}) should be slightly
modified to account for all guests arriving at a given node $n$:
\begin{eqnarray}
\!\!\dot{S}_{n} &=&-\beta _{n}S_{n}(I_{n}+\sum_{m\neq n}I_{mn})%
\hphantom{{}-\alpha _{n}I_{n}} -\sum_{m\neq n}\gamma
_{nm}^{S}S_{n}+\sum_{m\neq n}\delta _{nm}^{S}S_{nm}  \label{ODE:Sn:network}
\\
\!\!\dot{I}_{n} &=&\hphantom{-}\beta _{n}S_{n}(I_{n}+\sum_{m\neq
n}I_{mn})-\alpha _{n}I_{n} -\sum_{m\neq n}\gamma
_{nm}^{I}I_{n}+\sum_{m\neq n}\delta _{nm}^{I}I_{nm}  \label{ODE:In:network}
\\
\!\!\dot{S}_{mn} &=&-\beta _{n}S_{mn}(I_{n}+\sum_{m\neq n}I_{mn})%
\hphantom{{}-\alpha _{n}I_{mn}} +\gamma _{mn}^{S}S_{m}-\sum_{m\neq n}\delta
_{mn}^{S}S_{mn}  \label{ODE:Smn:network} \\
\!\!\dot{I}_{mn} &=&\hphantom{-}\beta _{n}S_{mn}(I_{n}+\sum_{m\neq
n}I_{mn})-\alpha _{n}I_{mn} +\gamma _{mn}^{I}I_{m}-\sum_{m\neq n}\delta
_{mn}^{I}I_{mn} . \label{ODE:Imn:network}
\end{eqnarray}%
Interaction between nodes is determined by $M\times M$ matrices $\gamma
_{mn}^{I,S}$, $\delta _{mn}^{I,S}$, and also matrices of guest populations $%
S_{mn}$ and $I_{mn}$. All these matrices have zero diagonal elements, and
they may also have zero elements if nodes $m$ and $n$ do not interact
directly.

The effective number of initial infectives for the network in the SIC
approximation can be calculated by the sum%
\begin{eqnarray}\label{eq:Ieff-network}
I_{0n}^{\mathrm{eff}} &\approx &\sum_{m\neq n}
 \int_{0}^{+\infty }
\left[C_{mn}^{(1)}I_{m}(t) +
C_{nm}^{(2)}S_{nm}(t)I_{m}(t)\right]e^{-\lambda _{n}t}\, \mathrm{d}t  \\ \nonumber
C_{mn}^{(1)} &=&\frac{\gamma _{mn}^{I}\beta _{n}N_{n}}{\beta
_{n}N_{n}+\delta _{mn}^{I}},\qquad C_{nm}^{(2)}\;{}={}\;\frac{\delta
_{nm}^{I}\beta _{m}}{\beta _{n}N_{n}-\alpha _{n}+\alpha _{m}+\delta _{nm}^{I}%
} .
\end{eqnarray}  

Consider, for example, an infinite 1D lattice of SIR centra where every
node $n $ interacts with its nearest neighbours $m=n-1$ and $m=n+1$. For
simplicity consider a network with identical centra: $\beta _{n}=\beta $, $%
\alpha _{n}=\alpha $, $N_{n}=N$ for $\forall n$. Also, let migration
parameters be identical for every node and for every population class: $%
\gamma _{mn}^{I,S}=\gamma $, $\delta _{mn}^{I,S}=\delta $ for $m=n\pm 1$ and
$\gamma _{mn}^{I,S}=0$, $\delta _{mn}^{I,S}=0$ otherwise. Then we obtain a
closed system of ODE
\begin{eqnarray}
\dot{S}_{n} &=&-\beta S_{n}(I_{n}+I_{n-1n}+I_{n+1n})-2\gamma S_{n}+\delta
S_{nn-1}+\delta S_{nn+1}  \label{1D:Sn} \\
\dot{I}_{n} &=&\hphantom{-}\beta S_{n}(I_{n}+I_{n-1n}+I_{n+1n})-\alpha
I_{n}-2\gamma I_{n}+\delta I_{nn-1}+\delta I_{nn+1}  \label{1D:In}
\end{eqnarray}
taking into account Eqs.~(\ref{ODE:Snm})--(\ref{ODE:Inm}) with $m=n\pm 1$%
\begin{eqnarray}
\!\!\!\dot{S}_{n\pm 1n} &=&-\beta S_{n\pm 1n}(I_{n}+I_{n-1n}+I_{n+1n})%
+\gamma S_{n\pm 1}-\delta S_{n\pm 1n}  \label{1D:Smn} \\
\!\!\!\dot{I}_{n\pm 1n} &=&\beta S_{n\pm
1n}(I_{n}+I_{n-1n}+I_{n+1n})-\alpha I_{n\pm 1n}+\gamma I_{n\pm
1}-\delta I_{n\pm 1n} .  \label{1D:Imn}
\end{eqnarray}

When parameters of nodes are identical we can introduce the universal
dimensionless time $t^{\prime }=\alpha t$ and dimensionless migration
parameters $\gamma ^{\prime }=\gamma /\alpha $, $\delta ^{\prime }=\delta
/\alpha $. Also define the shares of host and guest infectives $%
i_{n}=I_{n}/N $, $i_{mn}=I_{mn}/N$ and susceptibles $s_{n}=S_{n}/N$, $%
s_{mn}=S_{mn}/N$. Then we can rewrite the equations in terms of
dimensionless variables, omitting primes:
\begin{eqnarray}
\!\!\!\dot{s}_{n} &=&-\rho s_{n}(i_{n}+i_{n-1n}+i_{n+1n})\hphantom{{}-i_{n}}%
-2\gamma s_{n}+\delta s_{nn-1}+\delta s_{nn+1}  \label{1D:sn} \\
\!\!\!\dot{i}_{n} &=&\hphantom{-}\rho s_{n}(i_{n}+i_{n-1n}+i_{n+1n})-i_{n}-2\gamma
i_{n}+\delta i_{nn-1}+\delta i_{nn+1}  \label{1D:in} \\
\!\!\!\dot{s}_{n\pm 1n} &=&-\rho s_{n\pm 1n}(i_{n}+i_{n-1n}+i_{n+1n})%
\hphantom{{}-i_{n\pm 1n}}\quad {}+\gamma s_{n\pm 1}-\delta s_{n\pm 1n}
\label{1D:smn} \\
\!\!\!\dot{i}_{n\pm 1n} &=&\hphantom{-}\rho s_{n\pm
1n}(i_{n}+i_{n-1n}+i_{n+1n})-i_{n\pm 1n}\quad {}+\gamma i_{n\pm 1}-\delta
i_{n\pm 1n} . \label{1D:imn}
\end{eqnarray}

We search for the travelling wave in the form%
\begin{eqnarray}
s_{n}(t) &=&s^{\mathrm{tr}}(t-Tn),\ i_{n}(t)=i^{\mathrm{tr}}(t-Tn)
\label{eq:tr1} \\
s_{n\pm 1n}(t) &=&s_{\pm }^{\mathrm{tr}}(t-Tn),\ i_{n\pm 1n}(t)=i_{\pm }^{%
\mathrm{tr}}(t-Tn)  \label{eq:tr2} \\
s_{nn\pm 1}(t) &=&s_{\mp }^{\mathrm{tr}}(t-T(n\pm 1)),\ i_{nn\pm
1}(t)=i_{\mp }^{\mathrm{tr}}(t-T(n\pm 1))  \label{eq:tr3}
\end{eqnarray}%
where $T$ is the time lag between outbreaks in two neighbour nodes. Here $i^{%
\mathrm{tr}}(t),s^{\mathrm{tr}}(t)$ are the shares of hosts in node $n=0$; $%
i_{-}^{\mathrm{tr}}(t),s_{-}^{\mathrm{tr}}(t)$ are the shares of guests in
node $n=0$ arrived from node $n=-1$; $i_{+}^{\mathrm{tr}}(t),s_{+}^{\mathrm{%
tr}}(t)$ are the shares of guests in node $n=0$ arrived from node $n=+1$; $%
i_{-}^{\mathrm{tr}}(t+T),s_{-}^{\mathrm{tr}}(t+T)$ are the shares of guests
in node $n=1$ arrived from node $n=0$; $i_{+}^{\mathrm{tr}}(t-T),s_{+}^{%
\mathrm{tr}}(t-T)$ are the shares of guests in node $n=-1$ arrived from node
$n=0$.

Substituting (\ref{eq:tr1})--(\ref{eq:tr3}) into (\ref{1D:sn})--(\ref{1D:imn}%
) we obtain the system of ODEs%
\begin{eqnarray}
\dot{s}^{\mathrm{tr}} &=&-\rho s(i^{\mathrm{tr}}+i_{-}^{\mathrm{tr}}+i_{+}^{%
\mathrm{tr}})\hphantom{{}-i^{\mathrm{tr}}}-2\gamma s^{\mathrm{tr}}+\delta
s_{-}^{\mathrm{tr}}\left( t+T\right) +\delta s_{+}^{\mathrm{tr}}\left(
t-T\right)  \label{ode:tr:s} \\
\dot{i}^{\mathrm{tr}} &=&\hphantom{-}\rho s(i^{\mathrm{tr}}+i_{-}^{\mathrm{tr%
}}+i_{+}^{\mathrm{tr}})-i^{\mathrm{tr}}-2\gamma i^{\mathrm{tr}}+\delta
i_{-}^{\mathrm{tr}}\left( t+T\right) +\delta i_{+}^{\mathrm{tr}}\left(
t-T\right)  \label{ode:tr:i} \\
\dot{s}_{\pm }^{\mathrm{tr}} &=&-\rho s_{\pm }^{\mathrm{tr}}(i^{\mathrm{tr}%
}+i_{-}^{\mathrm{tr}}+i_{+}^{\mathrm{tr}})\hphantom{{}-i_{\pm
}^{\mathrm{tr}}}\quad {}+\gamma s^{\mathrm{tr}}(t\pm T)-\delta s_{\pm }^{%
\mathrm{tr}}  \label{ode:tr:s->} \\
\dot{i}_{\pm }^{\mathrm{tr}} &=&\hphantom{-}\rho s_{\pm }^{\mathrm{tr}}(i^{%
\mathrm{tr}}+i_{-}^{\mathrm{tr}}+i_{+}^{\mathrm{tr}})-i_{\pm }^{\mathrm{tr}%
}\quad {}+\gamma i^{\mathrm{tr}}(t\pm T)-\delta i_{\pm }^{\mathrm{tr}} .
\label{ode:tr:i->}
\end{eqnarray}%
A travelling wave should satisfy the initial conditions (dynamic equilibrium
in the absence of outbreak):%
\begin{equation}
s^{\mathrm{tr}}(-\infty )=1,\qquad i^{\mathrm{tr}}(-\infty )=0,\qquad s_{\pm
}^{\mathrm{tr}}(-\infty )=\bar{\varepsilon},\qquad i_{\pm }^{\mathrm{tr}%
}(-\infty )=0.  \label{tr:ini}
\end{equation}%
Here $\bar{\varepsilon}=\gamma /\left( \gamma +\delta \right) $ is the share
of guest susceptibles at equilibrium in the absence of an outbreak (see (\ref%
{S12:lim})), it will be used in the analysis of the influence of guest
susceptibles before the outbreak and slightly simplifies the equations.

When $t\rightarrow -\infty $ the travelling wave initially have exponential
growth. We substitute solution in the form%
\[
s^{\mathrm{tr}}(t)=1-\Delta s^{\mathrm{tr}}e^{\lambda t},i^{\mathrm{tr}%
}(t)=ie^{\lambda t},s_{\pm }^{\mathrm{tr}}(t)=\bar{\varepsilon}-\Delta
s_{\pm }^{\mathrm{tr}}e^{\lambda t},i_{\pm }^{\mathrm{tr}}(t)=i_{\pm
}e^{\lambda t}
\]
into (\ref{ode:tr:s})--(\ref{ode:tr:i->})\ and linearize the equations with
respect to $\Delta s^{\mathrm{tr}},i^{\mathrm{tr}},\Delta s_{\pm }^{\mathrm{%
tr}},i_{\pm }^{\mathrm{tr}}$.

The values $i,i_{-},i_{+}$ satisfy the following system of linear algebraic
equations%
\begin{eqnarray}
\lambda i &=&\rho (i+i_{-}+i_{+})-i-2\gamma i+\delta i_{-}e^{\lambda
T}+\delta i_{+}e^{-\lambda T}  \label{lambda:i} \\
\lambda i_{\pm } &=&\rho \bar{\varepsilon}(i+i_{-}+i_{+})-i_{\pm }\quad
{}+\gamma ie^{\pm \lambda T}-\delta i_{\pm }  \label{lambda:i->}
\end{eqnarray}%
which has the following characteristic equation
\begin{eqnarray*}
L \quad{=}\quad\lambda _{0}-\lambda -2\gamma +\frac{2\gamma \delta }{\lambda +1+\delta
}+\frac{2\bar{\varepsilon}\rho ^{2}}{\lambda +1+\delta -2\bar{\varepsilon}%
\rho }+\frac{2(\gamma +\bar{\varepsilon}\delta)\cosh \left( \lambda T\right) }{\lambda +1+\delta -2\bar{%
\varepsilon}\rho } \\
{}+\frac{2\,\bar{\varepsilon}\gamma \delta \cosh \left(
2\lambda T\right)}{\left( \lambda +1+\delta \right)
\left( \lambda +1+\delta -2\bar{\varepsilon}\rho \right) } \quad{=}\quad 0
\end{eqnarray*}%
where $\lambda _{0}=\rho -1$ is the initial growth rate of the limiting
solution in the dimensionless time $t^{\prime }=\alpha t$.

In the case $\bar{\varepsilon}=0$ we have%
\begin{equation}
L_{\bar{\varepsilon}=0}=\lambda _{0}-\lambda -\frac{2\gamma \left( \lambda
+1\right) }{\lambda +1+\delta }+\frac{2\gamma \left( \lambda _{0}+1\right) }{%
\lambda +1+\delta }\cosh \left( \lambda T\right) .  \label{eq:L:e=0}
\end{equation}%
By analogy with \cite{SKG11a} (Eq.~(42) there) we express these formulas in
terms of $\gamma $ and $\delta $%
\begin{equation}
L_{\mbox{\cite{SKG11a}}}=\lambda _{0}-\lambda -\frac{2\lambda \gamma }{%
\lambda +\gamma +\delta }\cosh \left( \lambda T\right) .  \label{eq:L:old}
\end{equation}
Last two formulas have some similarities but do not coincide exactly.

\begin{figure}[!h]
\centering \includegraphics[width=100mm]{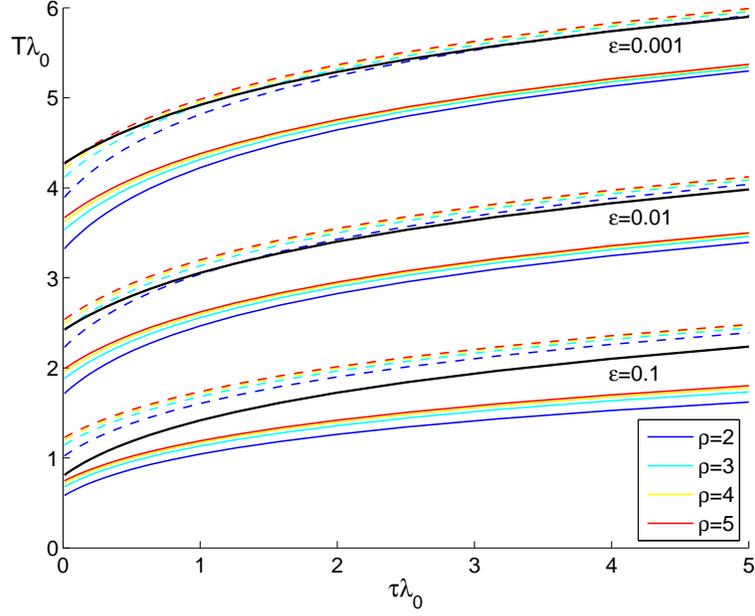}
\caption{Color solid curves depict dependence of the dimensionless time lag
(slowness) for traveling waves $\protect\lambda_0T$ on the characteristic
migration time $\protect\lambda_0\protect\tau$ for different $\protect%
\varepsilon\equiv\protect\gamma/(\protect\gamma{+}\protect\delta)$
(indicated near the curves) and reproduction number $\protect\rho$
(indicated in the legend). Colored dashed curve depict the same dependencies
in the case of an absence of guest susceptibles before the outbreak: $\bar%
\protect\varepsilon=0$. The black lines depict the similar curves taken from
work \protect\cite{SKG11a}; they are independent of $\protect\rho $.}
\label{fig:mvle}
\end{figure}

In accordance with the principle of linear spreading velocity
(LSV) (cf. \cite%
{Borsch90,Mollison91,Volpert,Mollison95,SKG11a}) we solve the system $L=0$, $%
\partial L/\partial \lambda $ with respect to $\lambda $ and $T$.

The results from numerical exercises are shown in Figure~\ref{fig:mvle} with
$\lambda _{0}T$ vs $\lambda _{0}\tau $ for different $\varepsilon =\gamma
/(\gamma +\delta )$ and $\rho $ (solid color lines) where $\lambda _{0}=\rho
-1$ is the initial growth rate for an individual SIR\ node in the SIC
approximation. Recall that in terms of dimensional parameters they are $%
(\rho -1)\alpha T$ and $(\rho -1)\frac{\alpha }{\gamma +\delta }$,
respectively. Here dashed color curves are plotted for the case $\bar{%
\varepsilon}=0$, i.e. neglecting guest susceptibles before the outbreak.
Also curves obtained in from Eq.~(\ref{eq:L:old}) are plotted by black
lines. We make the following observations:

1. Note that the functions $\lambda _{0}T(\lambda _{0}\tau )$ depend on $%
\rho $ but not to a large extent: the smaller coupling $\varepsilon $ ---
the smaller dependence. More discrepancy is observed for small $\tau $.

2. The curves for $\bar{\varepsilon}=0$ are very close to those obtained in
\cite{SKG11a}, especially for small $\varepsilon $.

3. Taking account of pre-outbreak guest susceptibles ($\bar{\varepsilon}\neq
0$) noticeably shortens the time lag, i.e. it accelerates the propagation of
the epidemic.

This indicates the importance of modelling guest populations separately.
Guest susceptibles have much higher probability of returning to the host
node than that for the simple migration process away from a home node.
Correspondingly those susceptibles being contaminated have a relatively high
probability of bringing back disease into their own host node, triggering an
outbreak there, and continuing the propagation of the overall epidemic.

\section{Conclusion}

We explore analytically and numerically the SIR
epidemic processes on a system of linked centra, and
investigate of the importance of population structure on the developing
dynamics of directly transmitted diseases. The deterministic model
of the migration process between two population centra demonstrates the
different r\^{o}le of susceptibles and infectives in the epidemic spread
across the population as a whole.

The careful consideration of this simplified deterministic model allows us
to derive the characteristic equation and hence to evaluate the speed of
epidemic propagation via the chain of similar nodes. The model demonstrates
that the epidemic speed is dependent on reproduction number $\rho$ but not to a large extent,
and this dependence declines to zero when the coupling vanishes. We also
show that the appearance of pre-outbreak susceptibles accelerates the
propagation of epidemic. The estimation of phenomenological parameters of
the deterministic model $\epsilon$ and $\tau$ requires disease specific
data. However, the study of pure migration (via questionnaire or transport
data) can be related to parameters $\gamma$ and $\delta$ (cf. (\ref{ken2a})).
We note, however, that these models serve as a hydrodynamic approximation
for a Markov process describing fluctuations in the discrete numbers of all
types of individuals involved. In the framework of stochastic models $\gamma$
and $\delta$ are treated as the transition rates, or probabilities of a
movements of a specific individual to a different center. A focus for future
work will be to estimate the influence of random fluctuations on epidemic
speed and discuss the more complicated situation of a network with several
routes of introduction of contamination in any particular population centre.


\end{document}